\documentclass[journal]{IEEEtran}

\usepackage{graphicx,epsf,epsfig}

\usepackage{subfigure}
\usepackage{color}
\usepackage{amsmath}
\usepackage{color}

\begin{document}
%
\title{\textcolor{black}{A quasi-analytical} model for energy-delay-reliability tradeoff studies \textcolor{black}{during write operations} in  perpendicular STT-RAM cell}
%
%
%

\author{Kamaram Munira, William H. Butler and Avik W. Ghosh

\thanks{K. Munira and A.W. Ghosh are with Charles L. Brown Department of Electrical and Computer Engineering, University of Virginia, Charlottesville,
VA 22903, USA. e-mail:munira@virginia.edu.}
\thanks{W.H. Butler is with the Department of Physics and Astronomy and Center for Materials for Information Technology, University of Alabama, Tuscaloosa,
AL 35487, USA.}
\thanks{Manuscript received \today.}}

\markboth{Journal of \LaTeX\ Class Files,~Vol.~6, No.~1, January~2007}%
{Shell \MakeLowercase{\textit{et al.}}: Bare Demo of IEEEtran.cls for Journals}
%

\maketitle

\begin{abstract}
One of the biggest challenges the current STT-RAM industry faces is maintaining a high thermal stability while trying to switch within a given voltage pulse and  energy cost.  In this paper, we present a physics based \textcolor{black}{analytical} model that uses a modified Simmons' tunneling expression to capture the spin dependent tunneling in a magnetic tunnel junction(MTJ). Coupled with an analytical derivation of the critical switching current based on the Landau-Lifshitz-Gilbert equation, and the write error rate derived from a solution to the Fokker-Planck equation, this model provides us a quick estimate of the energy-delay-reliability tradeoffs in perpendicular STTRAMs \textcolor{black}{due to thermal fluctuations}. In other words, the model provides a simple way to calculate the energy consumed during a write operation that ensures a certain error rate and delay time, while being numerically  far less intensive than a full-fledged stochastic calculation. We calculate the worst case energy consumption during anti-parallel (AP) to parallel (P) and P to AP switchings and quantify how increasing the anisotropy field $H_K$ and lowering the saturation magnetization $M_S$, can significantly reduce the energy consumption. \textcolor{black}{ A case study on how manufacturing variations of the MTJ cell can affect the energy consumption and delay is also reported}.

\end{abstract}

\begin{IEEEkeywords}
Spin polarized transport, Spin torque
\end{IEEEkeywords}

%
\IEEEpeerreviewmaketitle

\section{Introduction}

\IEEEPARstart{W}{ith} its fast write and read, small cell size, non-volatility and excellent endurance, Spin Transfer Torque-based Random Access Memory (STT-RAM) devices bear excellent potential of dominating the embedded and standalone memory world in the near future\textcolor{black}{\cite{spin,Wolf16112001}}. However, in order to compete with existing embedded technology, an STT-RAM bit cell should be able to meet the following requirements: (i) a tunneling magnetoresistance (TMR) needs to exceed 150\% for low-power read; (ii) thermal stability ($\Delta$) must be greater than 60 to achieve high static reliability; (iii) a scalable spin transfer torque switching current ($ J_c \le$ 1.5 x $10^6$ A/$cm^2$); (iv) a very low write error rate (WER) $\sim10^{-9}$ with a fast device write time $<$ 10ns; and (v) an overall low energy write ($<$1pJ), which corresponds to a low switching voltage \cite{ITRS}. Achieving such a diverse set of targets will require meticulous design and material engineering. 

\textcolor{black} {The reliability of the writing process in arrays  of STT-RAM cells is dependent on the following factors: (i) variation of the electronic, magnetic and geometrical parameters across the array (Fig. 1)\cite{5424242,augustine}, (ii) thermally activated initial angle, $\theta_0$, at the beginning of the writing process, (iii) thermal fluctuations during the writing process\cite{wang:034507} and (iv) voltage drop across the 1-transistor 1-MTJ cell\cite{5467394}. For this paper, we disregard the voltage drop across the transistor in (iv) as it it outside the scope of this work. We address the rest in our model.}

\begin{figure}[t!]
\epsfig{figure=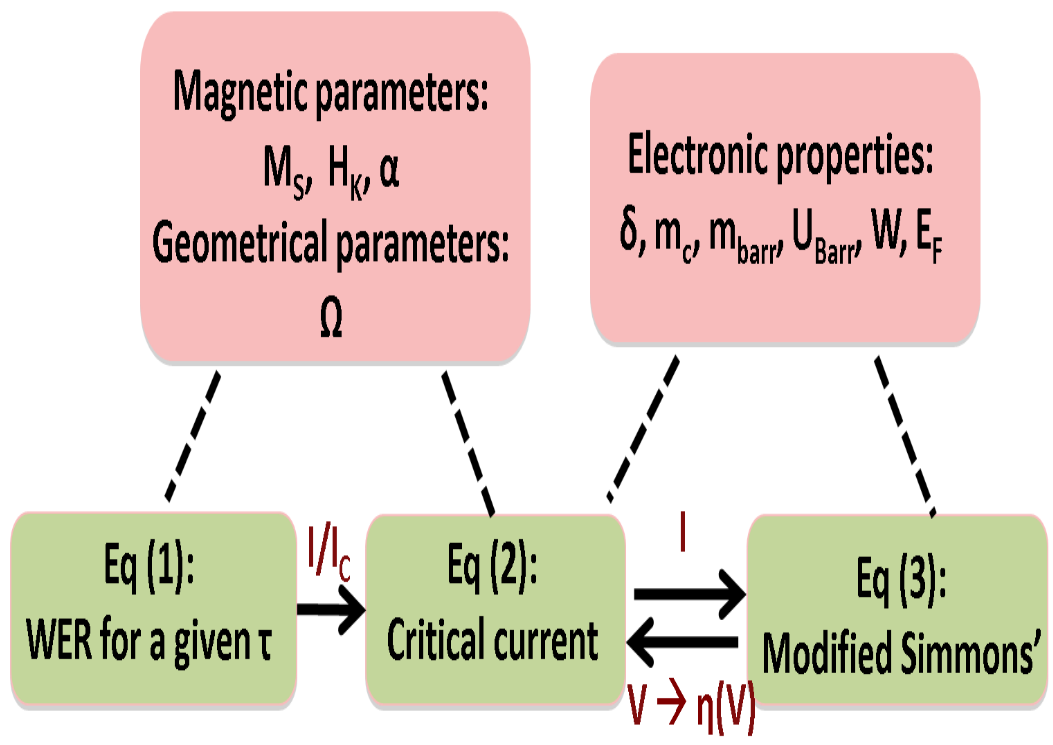,width=3.5in,height=1.3in}
\caption{Three coupled equation blocks allow us to compute the energy consumption for a write operation in a perpendicular STTRAM cell within a given switching time and error threshold. 
Equation block 1: Given a write-error-rate (WER) and delay time $\tau$, we use  the non-switching expression, Eq.1,  to extract the overdrive $ I/I_C$ ratio, where $I$ is the current applied and $I_C$ is the critical switching current. Equation block 2: gives us the critical switching current  $I_C$ using Eq. 2. From these two blocks, we get $I$. Equation block 3: Using a modified version of Simmons’ equation, Eq. 3,  we can then estimate the switching voltage $V$ and thence the switching energy $IV\tau$. The last two steps have to be solved self-consistently, as the switching voltage for a given WER depends on current $I_C$, which in turn depends on the voltage-dependent polarization $\eta(V)$, which further depends on the transport parameters $E_F$, $m_c$ and $\delta$.
} \label{band_structure}
\end{figure}

\textcolor{black}{(ii) and  (iii) are the dominant players when it comes to the writing process in STT-RAM, with (ii) being the most critical.} The free layer in the STT-RAM MTJ memory cell is responsible for data storage, and  must have enough stability against stochastic thermal switching to retain data for at least 10 years\cite{5467394}. In other words, the thermal stability, $\Delta$, needs to be greater than 60. While we desire the device to have high thermal stability, the critical current and the time needed to switch also increase proportionally. Higher thermal stability also contributes to longer switching tails, as the magnetization tends to be nearer to the stagnation points ($\theta=0~or~\pi$) in the potential landscape, without any external perturbation. Near the stagnation point, the torque proportional to $\sin\theta$ is too small to initiate switching. Therefore a thermally stable free layer would require a relatively high current to have very low error rate $10^{-9}$ for switching within 5-10 ns, resulting in costlier write operations. 

\textcolor{black}{While full-fledged stochastic material and circuit simulations can help narrow down the target compositions, they are time-consuming. Therefore, it is important to develop in parallel a fast, quasi-analytical physics based model to explore the multi-parameters optimization problem involved. This is the aim of the paper. }

In this paper, we introduce a \textcolor{black}{quasi-analytical} model  (Fig 1) that can be used to calculate the current and energy consumed during a write operation with an  acceptable WER for a given pulse width, $\tau$, in a perpendicular STT-RAM cell. The calculation is done in three blocks: (1) Given a write-error-rate and switching time, we use  an expression for switching errors from \cite{PNSW} to extract the $I/I_C$ ratio, where $I$ is the current applied and $I_C$ is the critical current. (2) The critical switching current  $I_C$ is calculated using the Landau-Lifschitz-Gilbert (LLG) equation to yield the applied current $I$. This expression also includes additional effects due to voltage-dependent polarization responsible for observed switching asymmetries\cite{Datta}. (3) Using a modified version of Simmons’ tunneling current equation, thats includes the crucial pre-factors responsible for TMR, we can then estimate the switching voltage and thence the switching energy cost. 
The last two steps are coupled and need to be solved self-consistently, as the switching voltage depends on critical current, which depends on polarization, which in turn  depends back on voltage. The energy consumed during the write operation can then be estimated from the switching current, voltage and time as $IV\tau$.

Coupling all three equations -- the WER expression, the modified Simmons tunneling equation  and the modified critical current
equation\cite{PhysRevB.62.570}  including the role of voltage-dependent polarization  are non-standard in the literature. Note that this paper focuses on perpendicular materials, for which the WER expression has
been explicitly worked out and calibrated against stochastic LLG solutions\cite{PNSW}. In future papers,
we will extend this study to in-plane materials as well as to double barrier STT-RAMs.  Section II describes the worst case \textcolor{black}{current needed for perpendicular STT-RAM}, based on solving the Fokker-Planck equation. In section III, we present the Simmons' equation modified to include spin dependent tunneling. The modified equations  are used to extract electronic parameters from published CoFeB/MgO/CoFeB MTJ current-voltage graphs. Using the parameters extracted from the fit in section III, the worst case energy consumption is thereafter calculated for a given $\Delta$ and error rate in section IV. \textcolor{black}{The effect of manufacturing variations on energy consumption is also studied.}

\section{\textcolor{black}{Current needed for worst case writing due to thermal fluctuations}}

\begin{figure}[t]
\centerline{\epsfig{figure=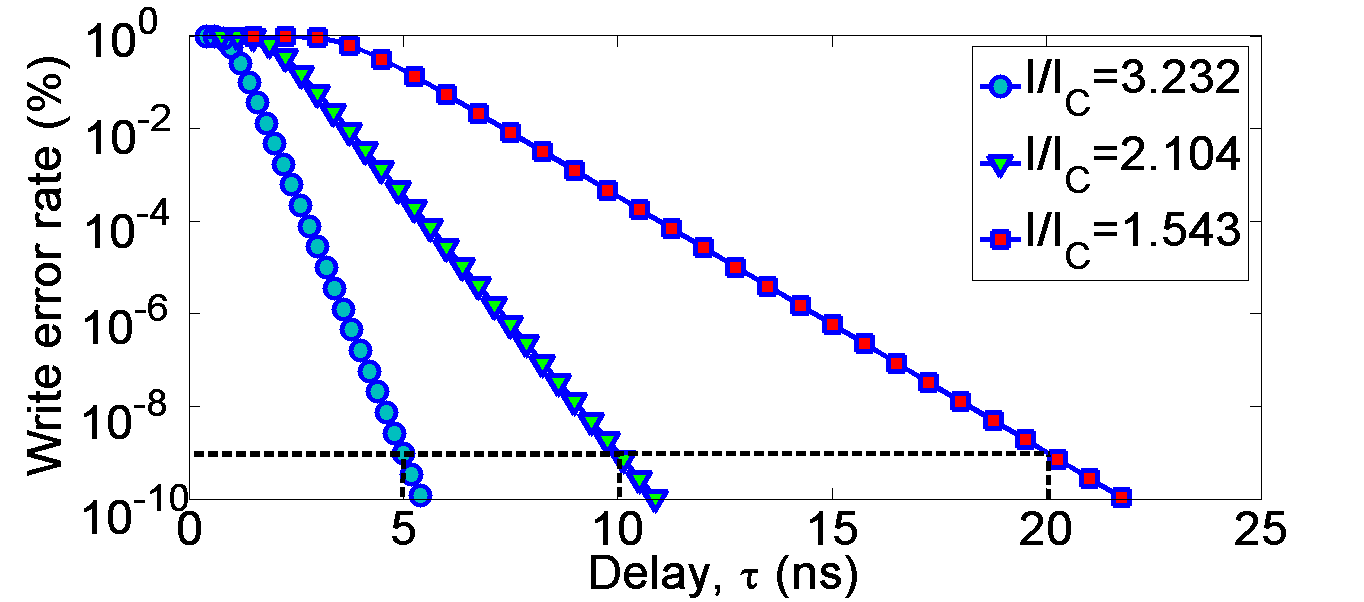,width=3.4in,height=1.8in}}
\caption{Write error rate vs. delay time at room temperature for $\Delta$=60. To attain an error rate of $10^{-9}$ within given delay times 5ns, 10ns and 20ns , current pulses of 3.232$I_C$, 2.104$I_C$ and 1.543$I_C$ are needed. We use $H_K$ of 3.34kOe and $M_S$ of 1257.3 $emu/cc$ for this estimate \cite{Ikeda:perpendicularCoFeB}.} \label{band_structure}
\end{figure}
Given a desired write-error-rate (WER) and switching delay $\tau$, we can calculate the amount of current overdrive required relative to the critical switching current \cite{PNSW}. Numerically this amounts to introducing a stochastic Langevin thermal torque in LLG and then taking a time average over many runs. The equivalent probabilistic approach is to solve the Fokker-Planck equation for the probability density of each configuration and then average over the distribution to extract various moments. The Fokker-Planck equation can be solved analytically for the simple case of a perpendicular material with uniaxial anisotropy \textcolor{black}{in the macrospin approximation}. The corresponding 
probability of {\it{not switching}} within a given delay $\tau$, defining the WER, is  expressed as 
\begin{eqnarray}
WER(\tau) &=& 1-\exp\biggl[\frac{-{\pi^2}\Delta(i-1)/4}{ie^{2{\alpha\gamma H_K \tau} (i-1)/(1+\alpha^2)}-1}\biggr]\nonumber\\
i &=& I/I_C
\end{eqnarray}
where $I$ is the current required and $I_C$ is the critical switching current, $\alpha$ is the Gilbert damping parameter, $\gamma$ is the gyromagnetic ratio, and $\Delta={H_KM_S \Omega}/{2k_BT}$ is the thermal stability of the free layer. $\Omega$ is the volume of the free layer and  $M_S$ is the saturation magnetization. $H_K=H_K^C-4\pi M_S^2$ is the effective anisotropy field resulting from both crystalline anisotropy($H_K^C$) and demagnetization field. $H_KM_S\Omega/2$ is the energy barrier that separates the two magnetization directions, $\theta=0$ and $\pi$. \textcolor{black}{Eqn. 1 is valid for precessional switching where $i>>1$.}

Solving the LLG equation we get a critical current similar to the expression derived by J.Z. Sun\cite{PhysRevB.62.570}, but modified to include the proper switching asymmetry through the effective voltage dependent polarization(described in the next section), 
\begin{equation}
I_C={2\alpha e H_KM_S \Omega}/{\hbar\eta(V)}
\end{equation}
From the given $WER$ and $\tau$, we can get the current overdrive $i = I/I_C$. In conjunction with the critical current above, we can then extract the applied current $I$. Finally, using the modified version of Simmons’ equation described in the next section, we can then estimate the switching voltage. 

\section{Modified Simmons' equation for magnetic tunnel junction}

In a magnetic tunnel junction, there are two ferromagnetic (FM) electrodes with an insulator between them. Ignoring atomistic effects such as symmetry filtering\cite{symmfilter}, the one-band alignment problem across the magnetic tunnel junction can be specified with six parameters: (i) $U$ is the barrier offset between the contact and the insulator, (ii) $E_{F}$ is Fermi-energy relative to the conduction band minimum in each contact (assumed to be comprised of the same material), (iii) $\delta$ is the band splitting between majority and minority spin electrons and related to the contact polarization, (iv) W is the width of the insulator, while (v) $m_{c}$ and (vi) $m_{barr}$ are the effective electron masses in the contact and barrier respectively (Fig. 3). At thermodynamic equilibrium, the Fermi levels  at the two FM electrodes are aligned. However for a positive bias $qV$ applied to the free layer, the energy levels in the fixed FM are shifted up by $qV/2$ and the ones in the free FM are lowered by $qV/2$, separating the Fermi levels at the two electrodes by $qV$. The mismatch in the Fermi levels cause current to flow in the system through oxide tunneling.  In a free electron model for the electrodes, the longitudinal spin-polarized electron momentum at a given energy $E$ in each of the three regions can be written as follows. Fixed FM: $k_{\uparrow, \downarrow}(V)=\sqrt{(2m_c[E_F- (\delta\mp\delta -qV)/2])}/\hbar$, Insulator: $\kappa=\sqrt{2m_{barr}U}/\hbar$ where $\beta=\kappa m_c/m_{barr}$, and Free FM: $k^{+}_{\uparrow,\downarrow}(V)=\sqrt{(2m_c[E_F- (\delta\mp\delta +qV)/2])}/\hbar$. The magnetization of the left contact is fixed, while the magnetization of the right contact is free to rotate and is defined by the relative angle, $\theta$. 

\begin{figure}
\centerline{\epsfig{figure=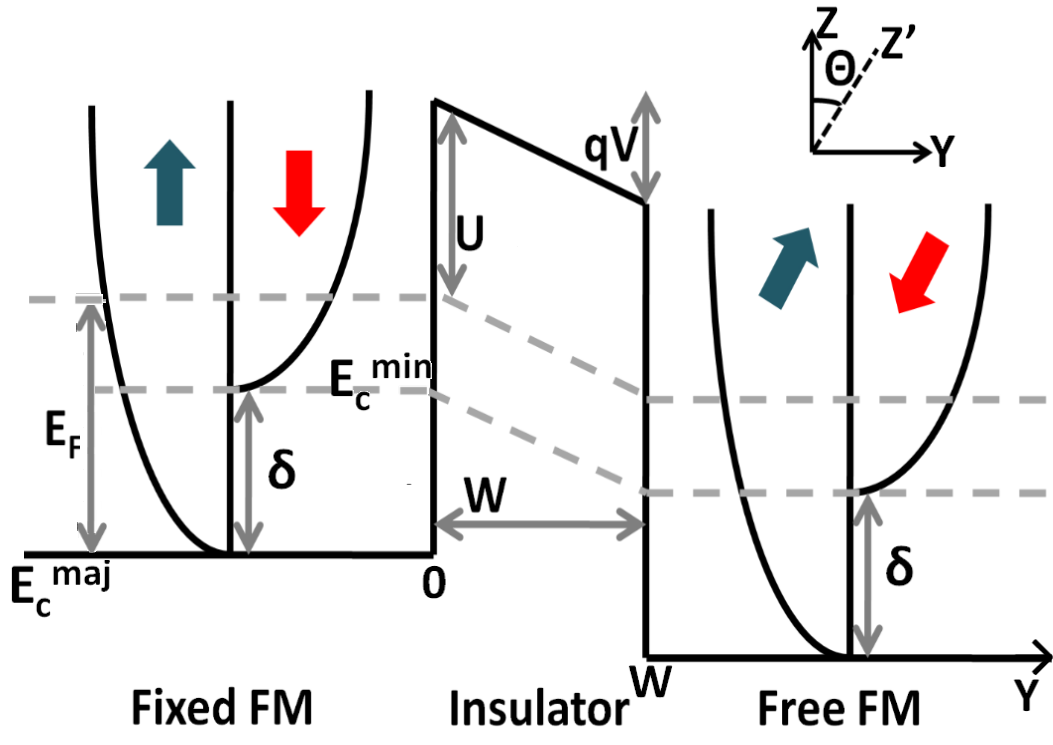,width=3in,height=1.8in}}
\caption{Schematic band alignment diagram of an MTJ. The bottoms of the $\uparrow $ and $\downarrow$  majority conduction electron bands in the FM contacts are labeled $E_C^{maj} and~ E_C^{min}$ respectively, and are separated by $\delta$ eV, while the insulating oxide introduces a tunnel barrier. $E_F$ is the Fermi-energy relative to the conduction band minimum, W  is the width of the insulating barrier and $U$ is  is the barrier offset between the contact and the insulator. 
 } \label{band_structure}
\end{figure}

 Using the methodology used in  \cite{PhysRevB.39.6995}, a weighted sum of the tunneling up and down spin electrons was obtained, with weighting factors given by their respective densities of states.  The transverse modes with momentum $k_{\parallel}$, and a linear potential drop in the insulator were included using the WKB approximation prescribed by Simmons in \cite{Simmons}. The total current density, J,  at a given voltage and angle $\theta$ in a magnetic tunnel junction is, 

\begin{align}
&J_{0}(V) =\displaystyle\frac{q^2}{4\pi^2\hbar W^2} \biggl[\bigl(U-
{qV}/{2}\bigr)e^{-\displaystyle \frac{2W}{\hbar} \sqrt{2m_c\bigl(U-{qV}/{2}}\bigr)}\biggl.\nonumber\\
&\hspace{2.6cm}-\biggr.\bigl(U+{qV}/{2}\bigr)e^{\displaystyle -\frac{2W}{\hbar} \sqrt{2m_c\bigl(U
+qV/2\bigr) }}\biggr]\nonumber\\
\nonumber\\
&J_{\sigma} =\displaystyle\frac{16k_\sigma\beta^2}{\beta^2+k_\sigma^2}
\biggl[\displaystyle\frac{k_\sigma^+\cos^2(\theta/2)}{\beta^2+k_\sigma^{+2}} + 
\frac{k_{\bar{\sigma}}^+\sin^2(\theta/2)}{\beta^2+k_{\bar{\sigma}}^{+2}}  \biggr],
 \begin{array}{lcr}
\sigma=\uparrow, \downarrow \\
\bar{\sigma}=\downarrow, \uparrow\end{array}\nonumber\\
\nonumber\\
&J(V,\theta)=\displaystyle[J_\uparrow(V,\theta)+J_\downarrow(V,\theta)]J_0(V)
\end{align}

From the spin current, we can calculate the torque exerted on the free magnetic layer \cite{PhysRevB.71.024411},
$\vec{T}(V) = (\hbar/2q)J_{avg}(V)\eta(V)[\hat{m}\times(\hat{m}\times\hat{m}_p)]$ where $J_{avg}(V)=[J(V,\theta=0)+J(V,\theta=\pi)]/2$. $\hat{m}$ and $\hat{m}_p$ represent
the unit vector orientations for the magnetizations in the free
and fixed layers respectively. When a positive bias is applied, polarized electrons flow from the  fixed FM contact to the free layer, causing the magnetization of the free layer to switch from anti-parallel to parallel. For negative bias, the electron flow removes magnetization from the right layer, causing it to rotate from P to AP. Notice that the torque exerted at the right, free ferromagnetic layer depends on the effective {\it{voltage-dependent}} polarization of the incident electrons from the left, fixed layer\cite{Datta}. This voltage dependence is critical to understanding the observed P to AP vs AP to P asymmetry in STTRAMs.
The effective polarization of the incident electrons at a given voltage is given by, 
\begin{align}
 \eta(V)=\displaystyle\frac{k_{\uparrow}-k_{\downarrow}}{k_{\uparrow}+k_{\downarrow}}
\end{align}
The voltage dependency on electron momentums, $k_\sigma$ and $k^+_\sigma$ are not explicitly mentioned in Eq. 3 and 4 for notational simplicity. Fig. 4 show results from the analytical model fitted with a published experiment. Eq. 3 was used to fit the resistance vs voltage data for in-plane CoFeB/MgO/CoFeB MTJ\cite{kuboto}. The parameters extracted from the model are listed in the figure caption.

\begin{figure}[t]
\centerline{\epsfig{figure=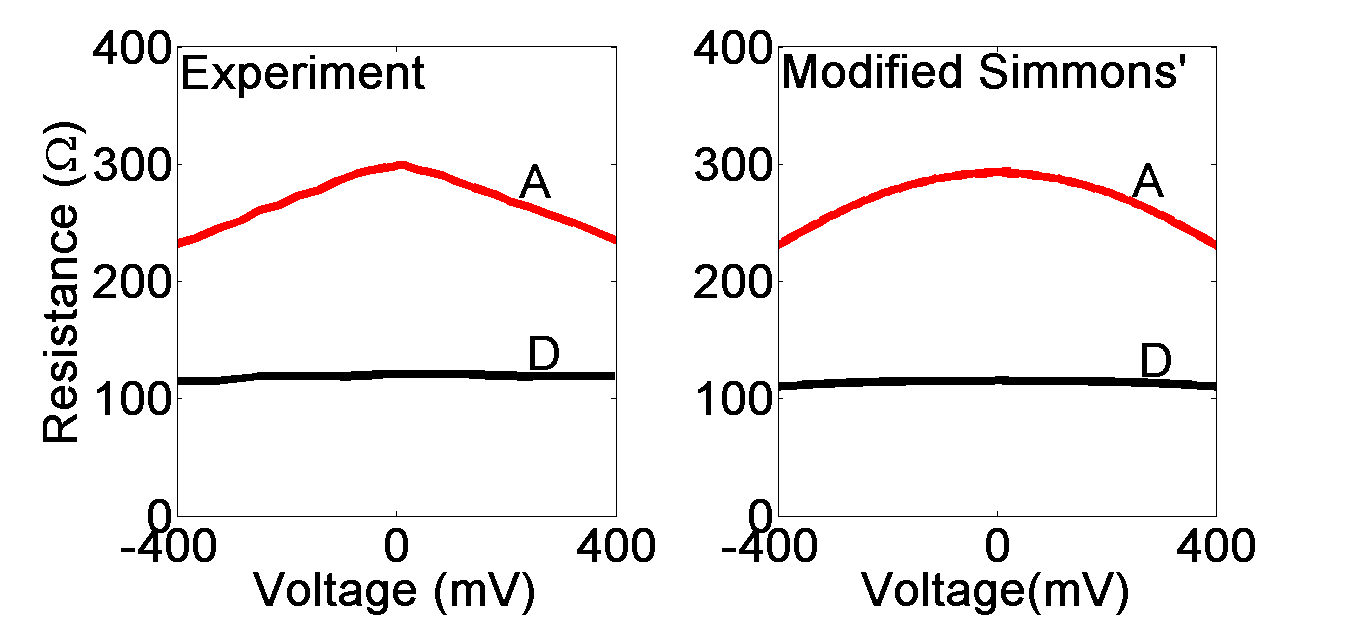,width=3.5in,height=1.8in}}
\caption{Parameter extraction from published CoFeB/MgO/CoFeB MTJ resistance-voltage curve\cite{kuboto} with modified Simmons' model. The parameters extracted from the model are $E_{F}$ =2.2eV, $U$= 1 eV, W = 1 nm,  $m_{c}$ = 0.3 $m_0$,  $m_{barr}$ = 0.18 $m_0$ and $\delta$ = 1.98 eV. Curve A is for anti-parallel mode while D for parallel mode. $m_0$ is the mass of an electron.}
\label{fig12}
\end{figure}

\section{\textcolor{black}{Worst case energy consumption and switching delay due to thermal fluctuations and manufacturing variations}}

Using the self-consistent analytical model illustrated in Fig. 1, we calculate the worst case energy consumption during a write operation at room temperature for $\Delta$=60 and various delays (Fig. 5) \textcolor{black}{due to thermal fluctuations}.  Energy consumption during P to AP switching is greater than that consumed during AP to P switching. This is because the effective polarization of the torque at negative voltage is less than that at positive voltage\cite{Datta,assymetry}. \textcolor{black}{However, for a magnetic tunnel junction with half-metallic contacts ($\delta$=3eV),  energy consumption during  AP to P switching is equal to that during P to AP switching because the effective polarization of the torque at negative voltage is same as that at positive voltage (Fig. 6). }

\begin{figure}[t]
\subfigure{\includegraphics[width=3.5in,height=1.8in]{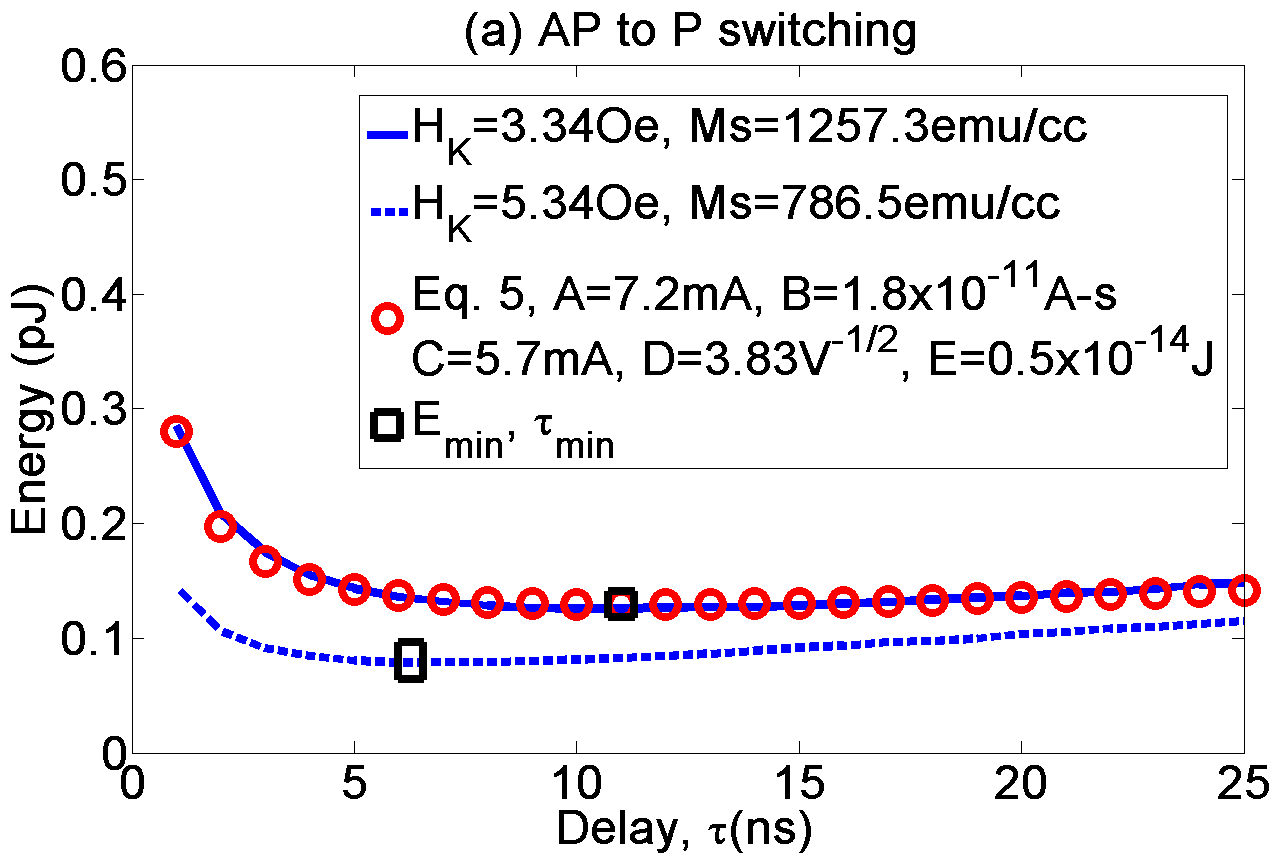}}
\subfigure{\includegraphics[width=3.5in,height=1.8in]{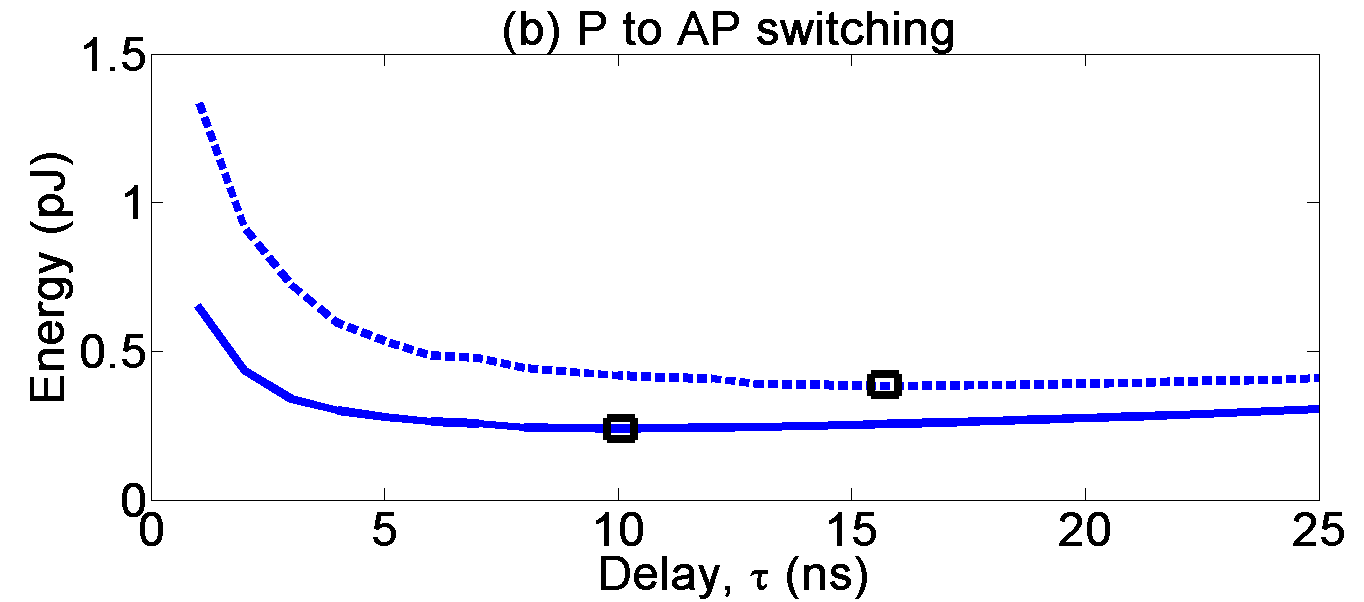}}\\
\caption{Worst case energy consumption during a write operation at room temperature for $\Delta$=60 and error rate of $10^{-9}$.  Energy consumption during (a) AP to P switching is less than that consumed during (b) P to AP switching. This is because the effective polarization of the torque at negative voltage is less than that at positive voltage. Energy consumed is the product of current, voltage applied and delay ($IV\tau$). At lower delay, the current, and hence energy, decreases logarithmically with delay. At higher delay, the energy consumption increases linearly with delay.} \label{band_structure}
\end{figure}

 Energy consumed is the product of current, voltage applied and delay ($IV\tau$). \textcolor{black}{The voltage drop considered here is the one across the MTJ. The additional voltage drop in the transistor which usually accompanies the STT-RAM cell is disregarded in this paper.} When the applied current is greater than the critical current, the switching is precessional and the switching delay of the free layer is inversely proportional  to the current, \textcolor{black}{$I\propto{I_C}+{I_C}(1+\alpha^2)ln(\pi/2\theta_0)/(\alpha \gamma H_K\tau)\propto (A\tau+B)$\cite{33spin, PNSW}, where $\theta_0$ is the initial angle of the switching free ferromagnet. Assuming that the current in the MTJ grows exponentially with applied voltage (Eq. 3), $V\propto [ln(I/C)/D]^2$. The energy consumption with respect to delay is }
\begin{align}
Energy \propto IV\tau\propto (A\tau+B)\biggl[ \frac{ln(\frac{A}{C}+\frac{B}{C\tau})}{D}\biggr]^2+E
\end{align}

\hspace{-4mm}\textcolor{black}{where A(=$I_C$), B(=${I_C}(1+\alpha^2)ln(\pi/2\theta_0)/(\alpha \gamma H_K)$), C(=$Area\cdot{q^2}/{4\pi^2\hbar W^2}$), D(=${2W}\sqrt{2m_c}/{\hbar} $) and E are constants. The units of constants A, B, C and E are Amp, Amp-sec, Amp, $V^{-1/2}$ and J respectively}. At lower delay, the  energy decreases logarithmically with delay as $ [ln(I/C)/D]^2$  is the dominant term. At higher delay, the energy consumption increases linearly with delay  as $A\tau$ is the dominant term. There is thus an energy minimum, $E_{min}$, corresponding to an optimal delay value, $\tau_{min}$. \textcolor{black}{ Fig.5 shows a fit of the numerical results with Eq. 5. The constants to fit the energy-delay curve for AP to P switching at $H_K=3.34kOe$ were: A=7.2mA, B=1.8x$10^{-11}$A-s, C=5.7mA, D=3.83$V^{-1/2}$ and D=0.5x$10^{-14}$J.}

 In addition, keeping $\Delta$ at 60, if $H_K$ is increased from 3.34 kOe to 5.34 kOe, and $M_S$ adjusted accordingly from 1257.3 emu/cc to 786.5 emu/cc, the  $E_{min}$ is shifted from  0.1252pJ at 10.5ns to 0.0783pJ at 6.6ns for AP to P switching. Similarly for P to AP switching, the energy minimum shifts lower from 0.3829pJ at 16.25ns to 0.2393pJ at 10ns. In other words, the  $E_{min}$ can be reduced quite significantly by increasing the anisotropy field, $H_K$ and decreasing the saturation magnetization, $M_S$ accordingly, while maintaining thermal stability. The duration of the voltage pulse needed to achieve the switching with $10^{-9}$ write error rate,  $\tau_{min}$,  also decreases with higher anisotropy field (Fig. 7).

\begin{figure}[t]
\centerline{\epsfig{figure=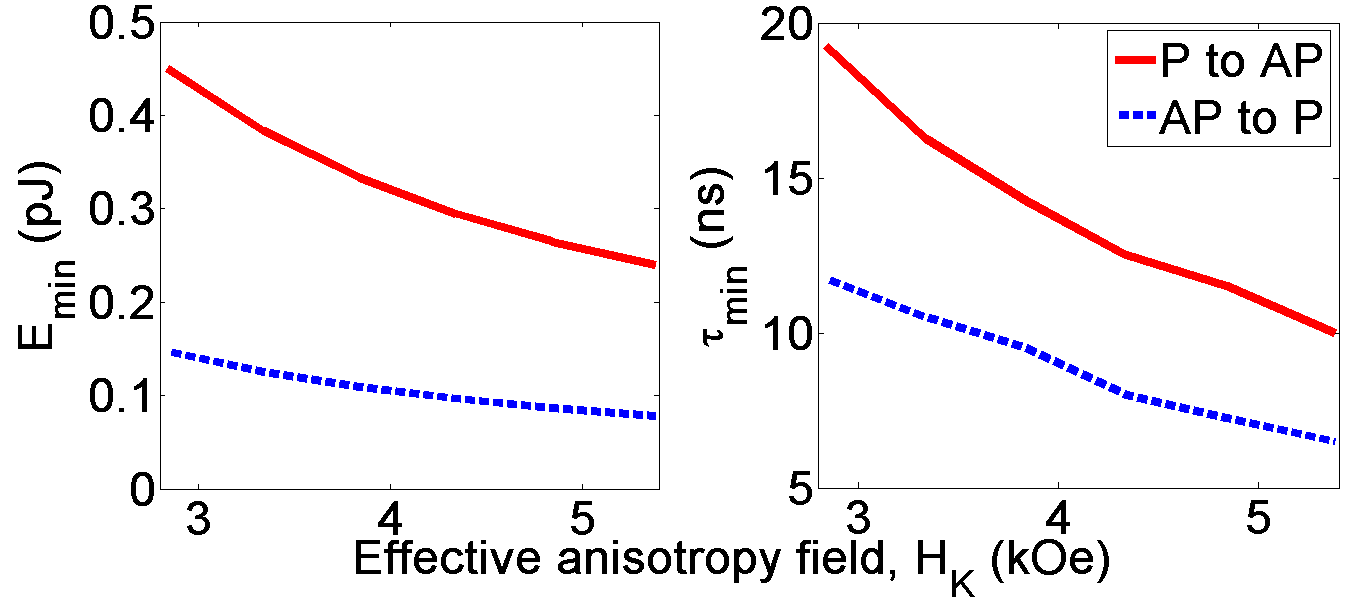,width=3.5in,height=1.8in}}
\caption{\textcolor{black}{Keeping volume and thermal stability of the free layer the same ($\Delta$=60), the minimum energy consumption can be reduced by increased the anisotropy field, $H_K$ and decreasing the saturation magnetization, $M_S$ accordingly. The duration of the voltage pulse needed to achieve the switching with $10^{-9}$ error rate also decreases with higher anisotropy field $\alpha$=0.02.} }\label{band_structure}
\end{figure}

\begin{figure}[t]
\centerline{\epsfig{figure=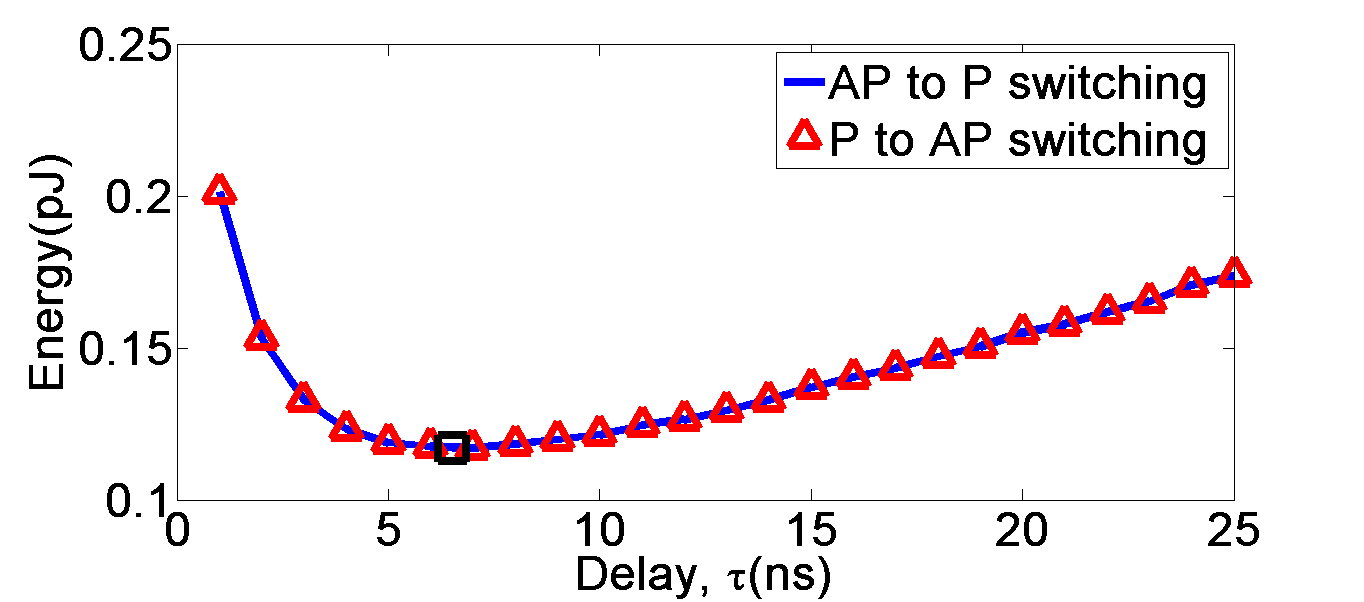,width=3.5in,height=1.8in}}
\caption{\textcolor{black}{Worst case energy consumption during a write operation at room temperature for $\Delta$=60 and error rate of $10^{-9}$ for a magnetic tunnel junction with half-metallic contacts ($\delta$=3eV).  Energy consumption during  AP to P switching is equal to that during P to AP switching because the effective polarization of the torque at negative voltage is same as that at positive voltage.  $H_K=5.34kOe$ and $M_S$= 786.5$emu/cc$.}
}  \label{band_structure}
\end{figure}

\textcolor{black}{ All the above simulation results (Figs. 5-7) only considered  thermally activated initial angles, $\theta_0$, at the beginning of the writing process and thermal fluctuations during the writing process in a given MTJ cell with a fixed dimension, electronic and magnetic parameters. However, individual MTJ cells in STT-RAM arrays will not all have the same physical dimension or parameters because of manufacturing variations. The key parameters, $H_K$, $M_S$, $\alpha$, cross-sectional area of the MTJ,  thickness of the free layer (t), W and U can vary by 10\% from their mean value\cite{5424242}. The self-consistent analytical model illustrated in Fig. 1 can be used to evaluate the write performance against various intrinsic variabilities as show in Figs. 8 and 9. All the above parameters were varied individually by $\pm$10\% and their effect on the $E_{min}$ and  $\tau_{min}$ are reported. The error rate is fixed at $10^{-9}$. Differing $H_K$, $M_S$, t and cross-sectional area will affect $\Delta$. Increased $\Delta$ and $\alpha$ increases the critical current and energy consumption. Increasing barrier width and height  results in  more voltage being needed to get the desired current.  While variation in the thickness of the tunneling barrier has the most effect, cross-sectional area has the least. The worst-case energy consumption and delay in an STT-RAM array can be calculated by considering an MTJ cell with biggest variation (+10\%) in all the parameters above. The values are 1.03pJ at 5.25ns for P to AP writing. A detailed study of the manufacturing variations will be reported in a later paper. }

\begin{figure}[t]
\centerline{\epsfig{figure=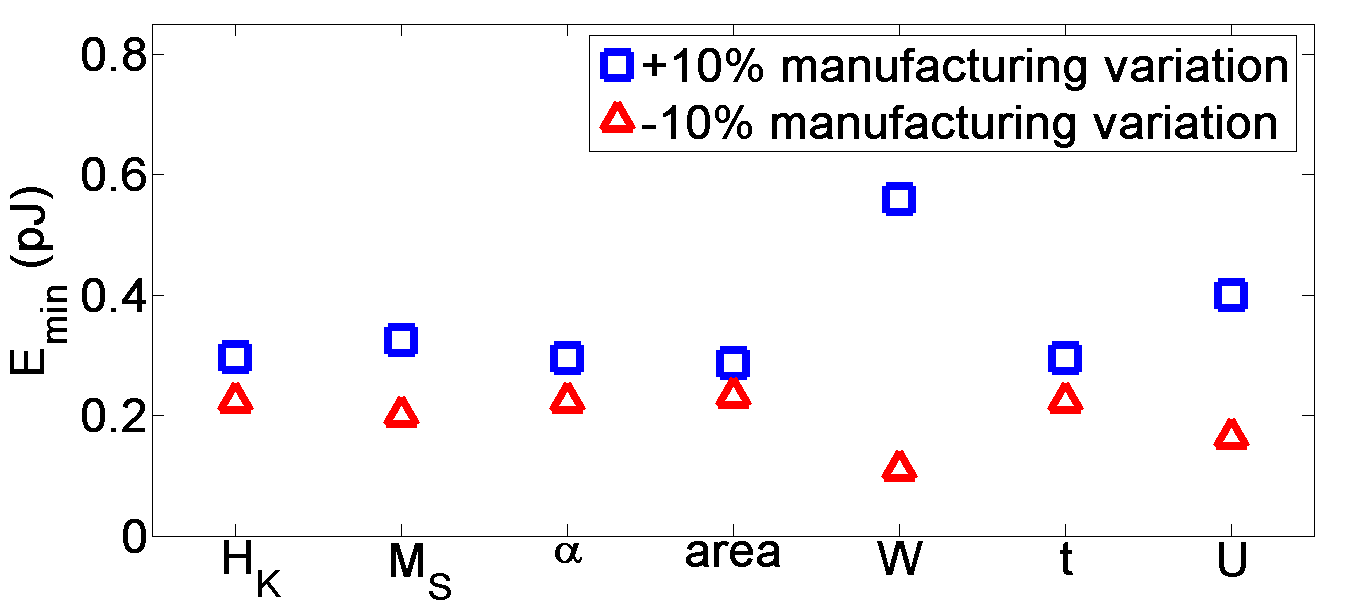,width=3.5in,height=1.8in}}
\caption{\textcolor{black}{Effect of manufacturing variations on minimum energy consumption, $E_{min}$, during a P to AP write operation with an error threshold of $10^{-9}$. The varying parameters are effective anisotropy field ($H_K$), saturation magnetization ($M_S$), Gilbert damping parameter ($\alpha$), cross-sectional area of the MTJ,  thickness of the free layer (t), width of the tunneling barrier (W), and ferromagnet-oxide interface (U).  While variation in the thickness of the tunneling barrier has the most effect on the delay, cross-sectional area has the least. Mean value of the MTJ parameters used are $H_K=5.34Oe$, $M_S=785.6$emu/cc, $\alpha=0.02$, radius of the free layer=20nm, t=2nm, W=1nm   and U=1eV. }} \label{band_structure}
\end{figure}

\section{Conclusion}
 In this paper, we present two models: (1) a modified Simmons' tunneling expression that precisely captures the spin dependent tunneling in a magnetic tunnel junction (Eq. 3) and (2) a corresponding model that approximately calculates the energy consumed during a write operation that ensures a certain error rate and delay time. The combination of the two models enables us to calculate the energy consumption without performing a full-fledged stochastic calculation and to evaluate ways to reduce the energy consumption during worst-case writing in STT-RAM. At lower delay, the energy consumption decreases logarithmically with delay while at higher delay,  it increases linearly with delay, creating a delay point where the energy consumption is minimum (Fig. 5). We  also quantified how increasing the anisotropy field $H_K$ and lowering the saturation magnetization $M_S$ while keeping $\Delta$ fixed, can significantly reduce the energy consumption and the switching delay. \textcolor{black}{The effect of the variations of geometric, magnetic and electronic properties in STT-RAM arrays is studied and it is seen that variation in the thickness of the oxide has the biggest role in determining the worst case energy consumption scenario.}

\begin{figure}[t]
\centerline{\epsfig{figure=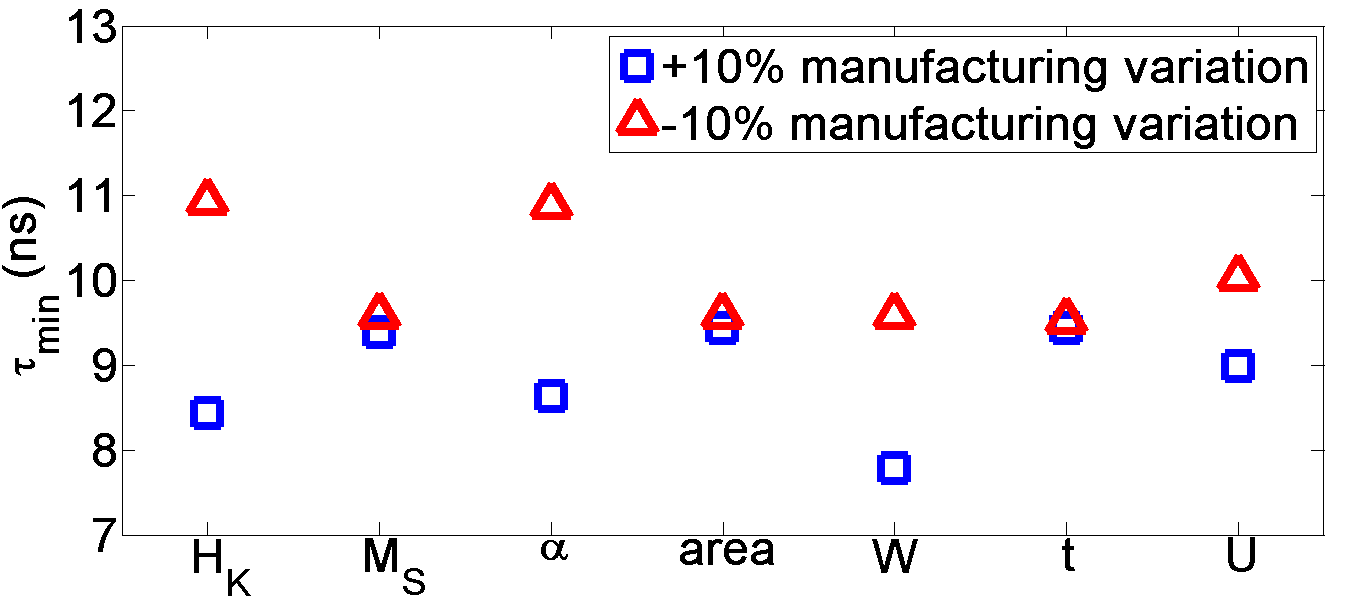,width=3.5in,height=1.8in}}
\caption{\textcolor{black}{Corresponding minimum switching delay during a P to AP write operation with an error threshold of $10^{-9}$. The varying parameters are effective anisotropy field ($H_K$), saturation magnetization ($M_S$), Gilbert damping parameter ($\alpha$), cross-sectional area of the MTJ,  thickness of the free layer (t), width of the tunneling barrier (W), and ferromagnet-oxide interface(U).}} \label{band_structure}
\end{figure}

\section*{Acknowledgment}
This work has been supported by DARPA grant HR0011-09-C-0023 and Samsung Electronics Corporation. We thank A. Nigam and M. Stan of University of Virginia and E. Chen of Samsung Electronics Corporation for useful discussions.

\bibliographystyle{IEEEtran}
\textcolor{black}{
\bibliography{circuit2}
}

\end{document}